# Comprehensive Study on Railway Communications Systems to Support Hyperloop

Hamid Amiriara
*Electrical Engineering*
Sharif University of Technology
*Tehran, Iran*
hamid.amiriara@sharif.edu

*Abstract*— Hyperloop is a sonic-speed train transporting passengers and freights in a vacuum tube without friction or air resistance. Two essential communications in such vehicles are central control connection and real-time dispatching, also an optional data connection for passengers is welcome. The high mobility of Hyperloop imposes a severe impact on the performance of wireless communication links. Therefore, designing a new wireless communication system is necessary to cope with the challenges. Motivated by the importance of doppler spreading, this paper focuses on the characterizations of the wireless channel in Hyperloop and then analyzes the performance degradation of the radio link. Afterward, a comprehensive overview of the present railway communication technologies as potential solutions for the Hyperloop project and a detailed discussion of their cons and pros are provided. It is shown that current communication technologies are not satisfactory for this scenario. Finally, this paper concludes the key points that need to be considered in future railway communications systems in order to overcome the Hyperloop communication challenges. So, we open up a new issue that Hyperloop communications require designing a novel method or improving the existing technologies or a combination of different techniques (some of these techniques are mentioned).

## I. Introduction

In recent years, the railway is overgrowing, and the vehicle's ultra-high-speed has drastically attracted worldwide attention. Now a day three types of high-speed rails are known worldwide. The first types of high-speed railway are the High speed Train (HST) with an average speed of 350 km/h. MAGnetic LEVitation (MAGLEV) trains are another high-speed railway that uses magnetic force (two magnet sets, one for pushing the train upward and the other for moving the train forward), with an average speed of 500 km/h. Traditional high-speed railways (HST and Maglev) generate air resistance commensurate to the square of the train speed and consume energy commensurate to the cube of the train speed. On the other hand, other types of high-speed railway are the Hyperloop which operates in an enclosed vacuum tube and can run at an ultra-high speed (333 m/s or 1200 km/h) (Fig. 1) [1].

This sonic-speed flying train in a vacuum tube does not meet traditional land transportation. According to the patent of Robert Goddard in 1945, it can be seen that the first investigation into Hyperloop began in 1909. However, Robert M. Salter's engineering articles from 1972 to 1978 made the vacuum train draw the public's attention, but the commercialization of the idea was impossible due to technological limitations and the vast costs at that time [2].

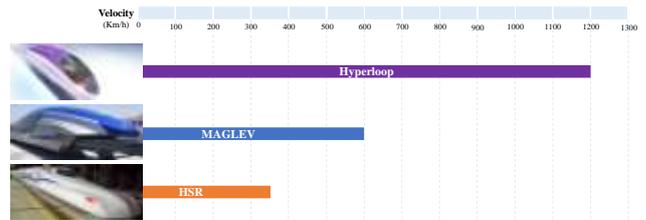

Fig. 1. Speed of different high-speed rails.

If the Hyperloop can travel with sonic speed at a vacuum tube by 2021, first, it will be deployed for freight and then for passengers. As the vehicle's speeds up, the safety of the train gains more attention. Land substructure (e.g., vacuum tunnel, track), movable substructure (e.g., train), and communication system are generally three key components in vehicle's safety. Thus, the communication system (e.g., the train control system), is the main part of the system and is considered the train nerve center. Reliable wireless communication is required for control of this vehicle (very reliable transmission at 48 kbps), which is low latency, high reliability, and capable of working at high-speed and the environment of a vacuum tube. Hyperloop safety also requires real-time monitoring inside and outside the train, so higher data rate vehicle-to-ground wireless communication (transmission at 100 Mbps) is required. In addition, passengers' internet access services in the Hyperloop may be required [3].

Wireless communication between the ground infrastructure and the moving vehicle (vehicle-to-ground) at sonic speed has not already been experienced in current land transport communications. In this study, we are looking to design a new vehicle-to-ground wireless communication system for ultra-high-speed rail technology in a vacuum tube.

## II. Performance Degradation of mobile radio applications Due to Doppler Spreading

We started by working on the performance analysis of orthogonal frequency division multiplexing (OFDM) systems for the sonic speed trains in terms of potential throughput. The orthogonality of the different subcarriers is destroyed due to the time variations of the channel during one OFDM symbol, which causes generating power leakage among the subcarriers. These destructive effects which are known as Inter-Carrier Interference (ICI) were carried out to realize the throughput degradation of OFDM systems due to doppler spreading, especially in sonic speed situations. The results show that for the Hyperloop scenario, the ICI decreases the performance of mobile radio applications, such as digital video broadcasting systems and IEEE 802.11a by 14.26 times and 22.5 times, respectively.

## A. System Model

Consider an OFDM system with $N$ subcarriers. Let $1/T$ be defined by the input symbol data rate; so the OFDM symbol interval is $NT$. The baseband transmitted signal can be modeled as

$$s(t) = \frac{1}{\sqrt{NT}} \sum_{k=1}^{N} s_k e^{j2\pi f_k t}, \quad 0 \leq t \leq N_T, \quad f_k = \frac{(k-1)}{NT} \quad (1)$$

where $s_k = \sqrt{P_T} d_k$ is the complex baseband symbol with the average power $P_T$ and $d_k$, is the normalized constellation point such as MPSK, MQAM, etc. $E[|d_k|^2] = 1$ [4].

Let $s(t)$ be transmitted over frequency selective channel with impulse response $h(t,\tau)$. Based on that sub-carrier is a narrow band so the channel of each sub-carrier can be modeled as a Rayleigh flat fading with impulse response $h_k(t,\tau)$. For the $k$th subchannel $h_k(t,\tau)$ is represented by

$$h_k(t,\tau) = a_k(t)\delta(\tau) \quad (2)$$

where the processes $a_k(t)$, $k = 1,\ldots,N$, are complex-valued jointly stationary and jointly Gaussian with zero means and variance function $E\{|a_k(t)|^2\} = \beta_k$. Then the baseband received signal can be expressed as

$$r(t) = h(t,\tau) * s(t) + n(t) = \frac{1}{\sqrt{NT}} \sum_{k=1}^{N} a_k(t) s_k e^{2\pi f_k t} + n(t) \quad (3)$$

where $*$ denotes convolution operation and $n(t)$ is additive whit Gaussian noise with spectral density $N_0$ watts/Hz.

The time-varying effects in the channel are sufficiently slow for Hyperloop radio fading channels, i.e., the coherence time is much larger than the OFDM symbol interval ($NT$). Based on the approach presented by Bello [5], the time-varying fading response of these slow fading channels can be expressed by two terms Taylor series as $a_k(t) = a_k(t_0) + \acute{a}_k(t_0)(t - t_0)$). Approximation of actual fading channel with the two-term Taylor series model is shown to be adequately providing that the maximum doppler frequency is less than 10 kHz (i.e., if a carrier frequency is 5GHz, the terminal must be moving at a speed of 2160km/h) [6]. This is larger than Hyperloop wireless fading channel for OFDM applications, therefore, this model is a sufficiently good approximation of the time variations encountered in Sonic-Speed Vacuum Train's channels. In this way, (3) can be written as

$$r(t) = \frac{1}{\sqrt{N_T}} \sum_{k=1}^{N} a_k(t_0) s_k e^{2\pi f_k t} + \frac{1}{\sqrt{N_T}} \sum_{k=1}^{N} \acute{a}_k(t_0)(t - t_0) s_k e^{2\pi f_k t} + n(t) \quad (4)$$

Each OFDM symbol interval of the received signal is passed through a parallel bank of correlators. The equation that describes the output of the $i$-th correlator is as follows:

$$\hat{d}_i = \frac{1}{\sqrt{NT}} \int_0^{N_T} r(t) e^{-j2\pi f_i t} dt \quad (5)$$

Substitution of (4) into (5) yields

$$\hat{d}_i = \frac{1}{\sqrt{NT}} \int_0^{N_T} \frac{1}{\sqrt{NT}} \sum_{k=1}^{N} a_k(t_0) s_k e^{2\pi (f_k - f_i)t} dt \rightarrow \text{Desired signal term}$$

$$+ \frac{1}{\sqrt{NT}} \int_0^{N_T} \frac{1}{\sqrt{NT}} \sum_{k=1}^{N} \acute{a}_k(t_0)(t - t_0) s_k e^{2\pi(f_k - f_i)t} dt \rightarrow \text{ICI term} \quad (6)$$

$$+ \frac{1}{\sqrt{NT}} \int_0^{N_T} n(t) e^{-j2\pi f_i t} dt \rightarrow \text{noise term}$$

Thus, the following equation is obtained:

$$\hat{d}_i = \underbrace{\sqrt{P_T} a_i(t_0) d_i}_{\text{Desired signal}} + \underbrace{\frac{\sqrt{P_T} N_T}{j2\pi} \sum_{\substack{k=1 \\ k \neq i}}^{N} \frac{\alpha_k \acute{a}_k(t_0) d_k}{(k-i)}}_{\text{ICI term}} + \underbrace{n_i}_{\text{noise}} \quad (7)$$

From (7), the mean of the desired signal power under the $i$-th sub-carrier can be computed by the following equation:

$$P_i = E\left\{\left|\sqrt{P_T} a_i(t_0) d_i\right|^2\right\} = \beta_i P_T \quad (8)$$

We can now proceed analogously to calculate the power of this interference (ICI). It can be shown that the first-order derivative process $\acute{a}_k(t)$ is a zero-mean complex Gaussian valued with a covariance $E\{|\acute{a}_k(t)|^2\} = 2\pi^2 \beta_k F_d^2$ [7]. (Maximum) doppler bandwidth is denoted ($F_d$) and defined as $F_d = v/\lambda$, for the train with speed $v$ and wavelength $\lambda$. Now we can derive the power of the interference (ICI) over the $i$-th sub-carrier as

$$I_i = \left(\frac{\sqrt{P_T} N_T}{2\pi}\right)^2 \sum_{\substack{k=1 \\ k \neq i}}^{N} \frac{E\{|\acute{a}_k(t_0)|^2\} E\{|d_k|^2\}}{(k-i)^2} = \frac{(\sqrt{P_T} N_T v)^2}{2\lambda^2} \sum_{\substack{k=1 \\ k \neq i}}^{N} \frac{\beta_k}{(k-i)^2} \quad (9)$$

Providing that $\beta_k(t_0)$ is independent of $d_k$, and the $d_k$'s are i.i.d. with zero means. Hence, we can derive the ratio of the desired signal power ($P$) to the interference power ($I$) and the noise power ($P_N$) for the $i$-th sub-carrier. By considering (8) and (9) we get SINR as

$$SINR_i = \frac{P_i}{I_i + P_N} = \frac{\beta_i P_T}{P_T \frac{(N_T v)^2}{2\lambda^2} \sum_{\substack{k=1 \\ k \neq i}}^{N} \frac{\alpha_k \beta_k}{(k-i)^2} + N_0} \quad (10)$$

Finally, the throughput of an OFDM system is calculated by considering the Inter-Carrier Interference (ICI) and the noise over a frequency-selective fading channel.

$$C = \sum_{i=1}^{N} B_i \log\left(1 + \frac{\beta_i}{\frac{(N_T v)^2}{2\lambda^2} \sum_{\substack{k=1 \\ k \neq i}}^{N} \frac{\alpha_k \beta_k}{(k-i)^2} + \frac{N_0}{P_T}}\right) \text{ bps} \quad (11)$$

## B. Simulation Results

In this section, to illustrate the sensitivity of the OFDM system's throughput to doppler related inter-carrier interference a simulation was performed, especially in the Hyperloop scenario for the existing standards.

Consider Fig. 2, which depicts the throughput of the OFDM system (Eq. 11) versus signal to noise ratio ($SNR = P_T/N_0$). The system parameters are: frequency-selective channel with bandwidth $BW = 1/T = 1MHz$, carrier frequency $f_c = 5GHz$ under doppler spread $F_d = 5.5$ kHz, corresponding to a train speed of 1200km/hr. The graph suggests that, for any value of $N$, the ICI is the limiting factor for performance at high SNR regimes. Also, it indicated that at the higher SNR, the ICI causes a considerable deterioration of throughput as the number of subcarriers, $N$, increases.

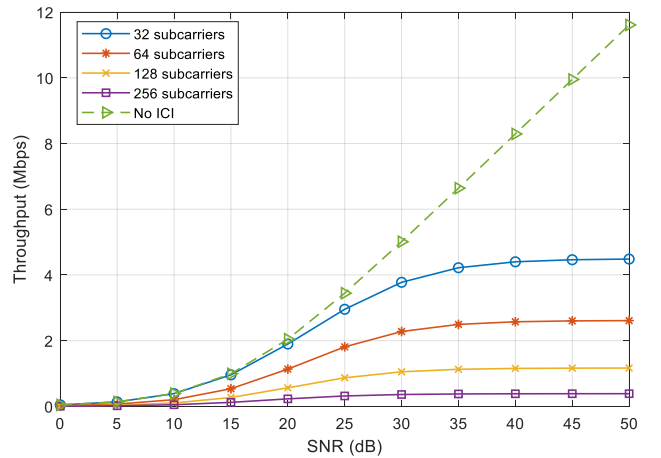

Fig. 2. Throughput of OFDM system (Eq. 11) versus SNR ($f_c = 5GHz$, $T = 1\mu s$, $v = 1200$ km/h).

TABLE I. DEPENDENCY OF OFDM SYSTEMS PERFORMANCE TO DOPPLER SPREADING IN DVB AND IEEE 802.11A SYSTEMS

| | Carrier Frequency ($GHz$) | Doppler Spread at 1200 km/h (kHz) | Subcarriers ($N$) | OFDM symbol ($NT$) (μsec) | Throughput at 1200km/h (MHz) | Throughput no doppler (MHz) |
|---|---|---|---|---|---|---|
| 802.11a | 5 | 5.9 | 64 | 4 | 212.2 | 189.7 |
| DVBCS2 | 4.8 | 5.78 | 2000 | 500 | 26.53 | 12.27 |

For a visual representation of the dependency of the OFDM system's performance on doppler spreading, the reader is referred to Table I which compares the throughput for two practical OFDM systems. One is a Digital Video Broadcasting (DVB) system (CS2 mode), and the other is an IEEE 802.11a WiFi system. Both systems are operating at 50dB SNR in a train traveling at a speed of 1200km/hr. As shown in Table I, the throughput degradation for the DVB-CS2 and IEEE 802.11a systems is different for the same channel under note that, their throughputs decreased by 14.26 times and 22.5 times, respectively, compared to the case of no-doppler spread.

### III. TRADITIONAL COMMUNICATION SYSTEMS IN RAILWAY SCENARIOS

We started by investigating whether existing wireless communication systems in railway scenarios can cover the requirement of the sonic-speed vehicle and vacuum tube Hyperloop. Traditional wireless communication technologies for railways can be categorized mainly into three classes: radio-based solutions, optical-based solutions, and satellite-based solutions. In the sequel, we introduce each of these traditional communication railways to determine the possibility of using each of these communication systems for application in Hyperloop.

#### A. Radio-based solutions

*Public Cellular Networks and Dedicated Infrastructures*
To enhance vehicle-to-ground data transmissions for high-speed trains, a large variety of technologies have been presented, namely Global System for Mobile Communications-Railway (GSM-R) [8], Advanced Long-Term Networks (LTE-A) [9], LTE for railway (LTE-R) [10] in the category of cellular Telecommunications solutions, and WiMAX on Rail [11], IEEE 802.11 [12] in the category of Dedicated Infrastructure. However, these mentioned technologies are capable of supporting data rates up to 100Mbps while moving at a maximum speed of 500km/h and are unable to meet Hyperloop's transport requirements.

*1. Public Cellular Networks*
In a particular case, the GSM-R can support vehicles with a maximum data rate of fewer than 200 kbps at a maximum speed of 500 km/h [4]. Evidently, this system cannot meet the requirement of data transmission in the Hyperloop project. In addition, LTE and LTE-R are well-known technologies that support wireless communications for vehicles traveling at a maximum speed of 350 km/h and 500 km/h, respectively. As noticed in the previous section the doppler frequency shift has a severe impact on the data rate for High-Speed railways, so it is not possible to achieve the maximum data rate at the sonic-speed vehicle with these technologies. Relying on the use of existing infrastructures, several publications have appeared in recent years to deploy Cellular-based technology for low costs railway communication. Nonetheless, because of the mobility of trains, Minimum capacity requires managing multiple cells. In addition, base stations (BTS) antennas are not oriented to cover the tracks, and they are often far from the tracks.

*2. Dedicated Infrastructure*
Much research has been done on technologies that can be deployed as a dedicated vehicle-to-ground infrastructure, in order to prepare communication on high-speed trains. Most of the previous studies paid considerable attention to WiMAX technology. However, this type of solution allows an entire control of the Quality of Service (QoS), especially in terms of throughputs and range. Reference [13] analyses and compares various aspects of WiMAX technology communication and claimed to provide throughput of up to 100Mbps at 350km/h. Of course, this technology is not yet widely used, and it wants to be used to provide passenger service in the Amtrak trains for the first time. The major drawback of dedicated infrastructure solutions for railway communication is the high costs in terms of operational expenditure and capital expenditure. So, there is a trade-off between costs and performance.

*3. Leaky Coaxial (LCX) Cables*
The LCX cable is a type of coaxial cable. The part of electric waves propagated through the cable radiate outside by periodic slots which are present in the outer conductor [14]. This generated circular symmetry electric field is confined to around the LCX; hence, it is very applicable for radio communication of moving objects in the tunnels, tracks, underground roads, etc.

In a recent paper by Chencheng et al. [15], the simulation of the LCX is conducted for a vacuum tube High-Speed flying train. The results show that the phase has uniform distribution along the tube, also the field distribution at the point away from the LCX is smooth, which can suppress doppler spreading. Therefore, the time-varying frequency-selective fading channel of a High-Speed flying train could be approximated as a stationary channel.

This communication method is advantageous because it can maintain a constant data transmission rate within one segment regardless of the position and speed of the vehicle, but it does not guarantee a data transmission rate of more than 10 Mbps due to the limit of leakage propagation. Its performance is also reduced due to handover (segment change) and a restriction on the segment length due to cable power loss. LCX can be guaranteed a reliable 1-10 Mbps data rate depending on the channel bandwidth 30-1000 MHz, but it requires the deployment of the LCX along the tunnel, which needs high installation and operation costs. In addition, the LCX system is non-scalable and no modification can be brought for the chosen operating frequency. There was a communication method based on LCX to support vehicles traveling at 350 km/h in HST or 600 km/h in MAGLEV. To our knowledge, there is no evidence to suggest that the system supports higher speeds (such as sonic speed or above).

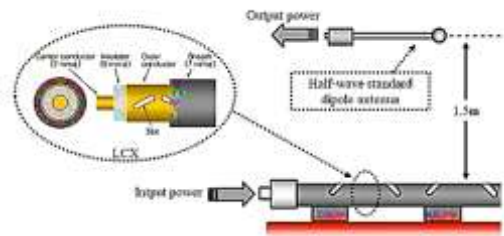

Fig. 3. Components of a leakage coaxial cable communication system.

## 4. Radio-over-Fibre (RoF)

Traditional cellular network throughputs significantly decrease in the case of high mobility, due to the transit from one BTS to another (frequent handovers or cell change). A solution is described in [16] that by reducing cell size, broadband connectivity can be achieved. Even though it leads to the use of a large number of BTS along the railway, the authors suggested deploying a Radio-over-Fiber (RoF)-based system. The RoF system is to transfer signal processing computing from BTS to a centralized control station, which leads to lower implementation costs and reduces the frequency of cell handovers. The RoF technology is used in a MAGLEV train Shanghai Transrapid, which throughputs of up to 4Mbps can be achieved in full-duplex at 3.5GHz and up to 16Mbps at 5.8GHz until 500 km/h [17], [18].

### B. Optic-based solution

Free-space optical communication (FSO) system uses free space light propagating to wirelessly transmit data for vehicle-to-ground communication. In this system, the incoming beam from Laser Diode (transmitter) passes through a cylindrical concave lens to form a beam with a width of the length of a train. The train can keep a communication link continuously to the land by projecting this wide fan-shaped beam to an Avalanche Photo Diode (receiver) where installed on a train (as shown in Fig. 4). Optical terminals have to be deployed at least every 400m along the track to keep a communication link continuously.

The experimental results in [19] show that a sufficient received Signal-to-Noise Ratio (SNR) could be obtained continuously even when the train is moving. This analysis led to the conclusion that the performance of the FSO system is expected to be a Giga-bit class communication for the high-speed train. However, an FSO-based solution can enable very high data rates until very high-speeds (For example, in a real test in Japan throughputs up to 100 Mbps were measured), but this solution has two major disadvantages. The first drawback of this method is the severe dependency of the performance on atmospheric conditions such as rain and fog. Another disadvantage of this solution is the heavy infrastructure installation which leads to very high expenditure and capital expenditure costs.

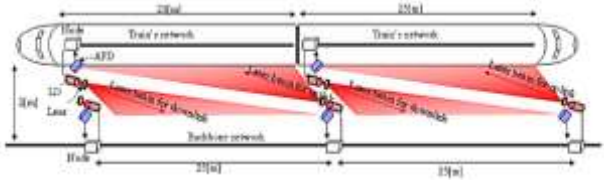

Fig. 4. A system model of FSO [19].

### C. Satellite-based solutions

Satellite-based communication also can be a solution to enable communication on ultra-high-speed trains.

The use of satellites leads to several advantages such as resistance to the high velocity of the vehicle, large geographical area coverage, well-adapted broadband, and low cost due to offer the advantage to be able to use the existing infrastructure and lack of installation of a dedicated infrastructure on the railway. However, many challenges will arise using satellites. Shadowing due to physical obstacles diminishes the line-of-sight broadband link. Also, the alignment of satellite antennas with their high spatial resolution beam and inherent train movement is not a trivial task. In addition, as illustrated in TABLE II., satellite links are expensive and they fall short of the requirement for the throughput.

TABLE II. THROUGHPUTS OF SATELLITE SOLUTIONS IN HIGH-SPEED TRAIN

| Name | Country | Throughput | Maximum speed |
|---|---|---|---|
| East Coast | England | 40Mbps (Claimed) | 225 km/h |
| Indian Railways | India | 9 Mbps | 320 km/h |
| NTV | Italy | 8 Mbps | 302 km/h |
| Thalys | Europe | 4 Mbps | 320 km/h |
| Temir Zholy | Kazakhstan | 2 Mbps | 250 km/h |

However, using the Ka-band with higher satellite capacity is a solution to increase performance and reduce the costs of the satellite solution. Nevertheless, studies on the investigation of cell changes and mobility effects of the Ka-band still have to be performed. This band is already used by Indian Railway, which reported throughputs up to 1.5Gbps. So, in the future, other kinds of satellites could be investigated in the Hyperloop project e.g., nanosatellites.

## IV. COMPARISON OF THE DIFFERENT TRADITIONAL COMMUNICATION SYSTEMS FOR USE IN HYPERLOOP

TABLE III summarizes the different traditional railway communication systems, in terms of throughputs, supported speed, tunnel usability, and some drawbacks and advantages. It is noteworthy that the direct link between the satellite and the train is essential in the satellite solution as well as the use of large antennas. So, this solution may not be conventional in the Hyperloop scenario as the vehicle is always moving inside the vacuum tunnel. In addition, the achievable data rate of these systems is limited (less than 10Mbps) and not very reliable, thus using this technology for any type of wireless connection required in the Hyperloop (Central control that requires very reliable communication and communication for real-time train dispatching 100 Mbps data transfer).

Now a day, many railway communications in worldwide use cellular solutions (GSM, GSM-R, LTE, and LTE-R). As mentioned in table 2, the main advantage of cellular communication has a new infrastructure to deploy and communicate by aggregating several public mobile network operators. But, the major drawback of this approach is the impossibility of QoS management, in terms of coverage, throughputs, latency, etc. The maximum rate supported by cellular communications systems is 10Mbps, which can be used for a vehicle traveling at a maximum speed of 500 km/h. As can be seen from Fig. 2, this method was tested on the sonic-speed conditions, and it is not possible to use this technology for both the required Hyperloop communications (very reliable 48 kbps connection, 100 Mbps data transfer).

Installing radio terminals, relying on Wi-Fi or WiMAX technologies, is the next solution. This system has high expenditure and capital expenditure costs due to requiring dedicated infrastructure. On the other hand, the throughput up to 100Mbps until 350km/h, can be obtained for open site scenarios. As a result, these technologies seem impossible for the Hyperloop scenario in that vehicle runs in a vacuum tunnel.

Another solution is connecting the different low-cost BTS with fiber optic. As mentioned in the previous section, this solution relies on a reduction of costs, a reduction of cell changes times, and an increase in throughputs. But this system support speeds up to 500km/h, therefore they are not usable in Hyperloop.

The optic-based solutions allow obtaining the best throughput and latency, and the speed does not impact the performance of optical-based systems, which is significant in the Hyperloop. The FSO can establish a Gbps level communication rate for the vehicle. However, optical-based solutions are dependent on atmospheric conditions. Furthermore, it requires a heavy infrastructure requirement along the track, and consequently, it is a very high cost.

The LCX-based solution is advantageous in that it can maintain a constant data transmission rate within one segment regardless of the position and speed of the vehicle. But a key limitation of this solution is that it is difficult to guarantee a data transmission speed of 10 Mbps or more due to the limit of leakage propagation, and it has performance degradation due to handover (segment change) and a restriction on the segment length due to cable power loss. It also requires cable deployment along the railway, which leads to high expenditure and capital expenditure costs. From the outcome of our investigation, it is possible to conclude that, this solution is appropriate to use for the Hyperloop-to-ground central control communication problem (very reliable communication at 48 kbps).

## V. Comparison of the Different Traditional Communication Systems for Use in Hyperloop

From the previous section, we have addressed only using LCX-based communication for the central control link but also dispatching link is essential in Hyperloop. For this link, we paid particular attention to cellular-based solutions. For ultra-high-speed moving such as the Hyperloop project, current cellular-based systems including the LTE and 4G are unable to provide high data rate services as much as low-speed users. So, the studies on the advanced cellular-based technologies for Hyperloop attract considerable attention.

In the last few years there has been a growing interest in Millimeter wave (mmWave) with vast spectrum resources at 28 GHz, 40 GHz, and 60 GHz, etc., as an attractive solution to provide high-data-rate transmission for future ultra-high-speed train communications [20], [21].

Several publications have appeared in recent years documenting the potentiality and feasibility of mmWave communications as a promising solution for railway networks. One of the first examples of a comprehensive design of a mmWave-based system is presented in [22], which demonstrated that a downlink throughput of 7Gbps can be achieved when the train's speed is up to 500km/h.

The literature on millimeter-wave communications shows a variety of approaches to deal with the high mobility of the train problems, including the large doppler spreads, fast fading channel, and frequent handover such as beam switching [23], antenna hopping [24], hybrid beamforming [25], and distributed antenna system (DAS) [26], etc.

As reported by [23] mmWave systems require proper beam alignment to achieve good performance making it difficult to apply to high-speed trains due to the need for frequent realignment. The authors investigated the optimal beam width that leveraged the position information from the train control system for efficient beam alignment and showed that a properly optimized system could achieve Giga-bit class throughput.

The most challenging issue in high-speed wireless communication like the Hyperloop project is estimating this extremely fast-fading channel. Jiao et al. [24] exploited delay correlation with sequential antenna-hopping (AH) and converts the rapid fading channels to a virtual slow-fading channel under high mobility.

Yaping et al. [25] refer that due to high channel correlation the line of sight (LOS) is an essential characteristic of high-speed railway propagation channels. For this reason, conventional multiple-input multiple-output (MIMO) such as VBLAST, and STBC is less effective under this scenario. Furthermore, it is hard to achieve multiple antennas gains with the high complexity of conventional MIMO. Spatial modulation (SM) is high spectral efficiency multiple antenna technology with low complexity that may be a potential technique for improving the performance of high-speed railways. A new hybrid SM beamforming scheme in mmWave frequency bands was proposed for the high-speed railway in reference [25]. Measured data in [27] imply the feasibility of SM for high-speed railways. The authors in [27] proved that the performance can be improved with an SM scheme system under high-speed wireless communication systems. Another type of solution is to use hybrid beamforming architecture, which consists of analog and digital domains [28]. In this work and related references, it was observed that digital precoding is performed to provide the necessary flexibility, and analog beamforming which formed the sharp RF beams is used to compensate for the severe path loss at higher frequency bands. The compromise between the complexity and performance of hybrid beamforming architecture, which jointly combines baseband

TABLE III. TRADITIONAL COMMUNICATION SYSTEMS FOR RAILWAY

| | Satellite | GSM-R | LTE-R | Radio terminals (IEEE 802.11-WiMax) | RoF | FSO | LCX |
|---|---|---|---|---|---|---|---|
| **Throughput** | >10Mbps | 172 kbps, | UL: 10Mbps, DL: 50Mbps, | >100 Mbps | 1 – 10Gbps | >10 Gbps | 1–10 Mbps |
| **Frequency** | Limited | UL: 876 – 880MHz, DL: 921 – 925MHz, | 450MHz, 800MHz, 1.4GHz, 1.8GHz | Variable | 3.5GHz, 5.8GHz | 830nm-1550 nm wavelength | Variable |
| **Channel bandwidth** | >20 MHz | 200 kHz | 1.4 – 20MHz | - | 10 – 100MHz | - | 30 – 1000MHz |
| **Mobility** | unlimited | <500km/h | <350km/h (LTE) <500km/h (LTE-R) | <350km/h | <500km/h | unlimited | unlimited |
| **Modulation multiplexing** | FSK-PSK | GMSK TDMA (SISO) | QPSK, 16-QAM, 64-QAM (SCFDMA, OFDM) (2×2 MIMO) | Variable | 16-QAM QPSK, (OFDM) | OOK, M-PPM and DPSK | Std. and OFDM |
| **OSS*/TS*** | OSS | OSS | OSS | OSS | OSS | OSS & TS | OSS & TS |
| **Advantages** | Available Infrastructure | No-infrastructure | No-infrastructure | Seamless communication | Seamless communication, Low cost | Seamless communication, High throughput | Seamless and reliable communication |
| **Drawbacks** | Limited throughput | Limited coverage, Limited throughput | Limited coverage, Limited throughput | High cost | High cost | Very high cost, Heavy infrastructure, Impact of atmospheric conditions | High cost, limited throughput due to limit of leakage propagation |

*Open Site Scenario
**Tube Scenario

precoding and RF beamforming, is discussed in [29]. [26] developed a distributed antenna system (DAS)-based mmWave communication system for HSTs to serve high-data-rate communication in high-speed railways. The authors introduced a system with moderate implementation complexity, based on the proposed network structure taking advantage of the distributed antenna system and millimeter-wave. Simulation results demonstrated that it is possible to provide Giga-bit class data services for high-speed railways.

In a recent paper by Qiu et al. [30], a structure was proposed that keeps the angle between radiation direction and movement direction as small as possible so the doppler shift could be regarded as a constant approximately. By combining two signals received from two sides of the train, the doppler spreading will turn into a real constant, which affects only the amplitude of received signals.

## VI. CONCLUSIONS AND FUTURE WORK

To meet future emerging ground and air vehicles networks targets, efficient vehicle-to-ground communications techniques are required, possibly with very reliable transmission at 48 kb/s for central control connections and transmission at 100 Mbps for real-time dispatching connections. To make this possible, based on the conducted study, we considered the railway communications techniques and developed the key points that need to be considered while formulating a throughput problem. Initially, we present a novel performance analysis of the OFDM systems for the sonic speed trains in terms of potential throughput. The Inter-Carrier Interference (ICI) destructive effect was carried out to realize the throughput degradation of OFDM systems due to doppler spreading especially in sonic speed situations. Our results describe for the first time the ICI decreased the performance of mobile radio applications, such as DVB-CS2 and IEEE 802.11a by 14.26 times and 22.5 times, respectively for the Hyperloop scenario. Then, we introduced some traditional communication systems in railway scenarios and highlighted the current research focuses and major challenges in these areas. However, to the author's best knowledge, very few publications are available in the literature that addresses the issue of wireless communication for the sonic-speed trains in a vacuum tube (Hyperloop). But, according to studies in the literature on a high-speed train, it seems that mmWave communications should be employed to deliver high data rates (achieving transmission at 100 Mbps for train dispatching) in Hyperloop services. As a next step, we provided vital features and entities involved in mmWave and railway communication with a detailed literature survey. It is found that, unlike the traditional high-speed railways, new technologies in mmWave communications such as hybrid beamforming, distributed antenna system, antenna hopping and beam switching, etc., should be used in the Hyperloop project. Current research efforts are also addressing the concepts that allow the doppler shift turns from a complex value into a real constant by combining two signals received from two sides of a flying train to reduce the influences only on the amplitude of received signals. Lastly, we found key points that needed to be considered while designing an efficient communications system to support sonic-speed vacuum trains and research directions for future evolution.